# LARGE PERIODIC TIME VARIATIONS OF TERMINATION SHOCK PARTICLES BETWEEN ~0.5 – 20 MeV AND 6-14 MeV ELECTRONS MEASURED BY THE CRS EXPERIMENT ON VOYAGER 2 AS IT CROSSED INTO THE HELIOSHEATH IN 2007: AN EXAMPLE OF FRESHLY ACCELERATED COSMIC RAYS?

W.R. Webber[1], A.C. Cummings[2], E.C. Stone[2], F.B. McDonald[3], D.S. Intriligator[4], P.R. Higbie[5], B. Heikkila[6] and N. Lal[6]

1. New Mexico State University, Department of Astronomy, Las Cruces, NM 88003, USA
2. California Institute of Technology, Space Radiation Laboratory, Pasadena, CA 91125, USA
3. University of Maryland, Institute of Physical Science and Technology, College Park, MD 20742, USA
4. Carmel Research Center, Space Plasma Laboratory, Santa Monica, CA 90406, USA
5. New Mexico State University. Physics Department, Las Cruces, NM 88003, USA
6. NASA/Goddard Space Flight Center, Greenbelt, MD 20771, USA

**ABSTRACT**

We have examined features in the structure of the heliosheath using the fine scale time variations of termination shock particles (TSP) between ~0.5 – 20 MeV and electrons between 2.5-14 MeV measured by the CRS instrument as the V2 spacecraft crossed the heliospheric termination shock in 2007. The very disturbed heliosheath at V2 is particularly noteworthy for strong periodic intensity variations of the TSP just after V2 crossed the termination shock (2007.66) reaching a maximum between 2007.75 and 2008.0. A series of 42/21 day periodicities was observed at V2 along with spectral changes of low energy TSP and the acceleration of 6-14 MeV electrons. Evidence is presented for the acceleration of TSP and electrons at the times of the 42/21 day periodicities just after V2 crossed the HTS. Spectra for TSP between 2-20 MeV and electrons between 2.5-14 MeV are derived for three time periods including the time of the HTS crossing. The energy spectra of TSP and electrons at these times of intensity peaks are very similar above ~3 MeV, with exponents of a power law spectrum between -3.0 and -3.6. The ratio of TSP intensities to electron intensities at the same energy is ~500. The electron intensity peaks and minima are generally out of phase with those of nuclei by ~1/2 of a 42 day cycle. These charge dependent intensity differences and the large periodic intensity changes could provide new clues as to a possible acceleration mechanism.

## Introduction

Four major populations of energetic particles above ~1 MeV are believed to exist in the heliosheath region: [1] Termination Shock Particles (TSP); [2] Anomalous Cosmic Rays (ACR); [3] Galactic Cosmic Rays (GCR); and [4] Galactic Electrons (Ge) (see e.g., Stone, et al., 2008). Of these populations only GCR exist in sufficient numbers to be well studied near the Earth. The other components all have intensities in the heliosheath which reach values $10^2$-$10^4$ times those which manage to reach the inner heliosphere near the Earth. In fact the TSP and ACR in the heliosheath behave much like "trapped particles" accelerated near to or just beyond the HTS or in the heliosheath itself, with the intensities building up to a level at which the production in the heliosheath equals the escape from both the inner boundary (HTS) and the outer boundary, the heliopause (HP).

These intensities and their fluctuations with time can be studied with a very high precision with the large area detectors which are part of the CRS instruments on V1 and V2 (Stone, et al., 1977). In this paper we describe some of these observations just after V2 crossed the HTS in 2007. The charged particle observations related to this crossing have been initially discussed by Stone, et al., 2008, and Decker, et al., 2008. The new studies at V2 reported here indicate a very turbulent and structured region near to and just beyond the HTS with strong ~42 day periodic intensity variations with sub-multiples of ~21 days, perhaps indicative of shocks, MIR's etc, propagating from the inner heliosphere into the heliosheath in route to the HP and beyond. The periodic structures just beyond the HTS on V2 may be an important source of accelerated TSP and electrons as evidenced by the periodic intensity increases. In what follows we will discuss the most significant of these intensity variations and what they imply.

## The Data- General Comments

In Figure 1 we show the rates (intensities) of >0.5 MeV H nuclei, representing TSP, and electrons from 6-14 MeV observed by the CRS instrument on V2. Also shown in this figure, for comparison, are the intensities of 27-42 MeV He nuclei as representative of ACR. This figure is expanded and is linear in the intensity dimension. This is in contrast to typical logarithmic plots which may extend several decades in intensity. The two time intervals labeled ① and ② in this figure bracket the time period when the largest periodicities in the intensity of TSP are observed. Time period ② will be discussed in a later paper.

It is seen from the Figure that the largest TSP intensities are observed directly beyond the HTS, crossed at 2007.66, where the dominant time variations are massive periodic variations. In the remaining time that V2 is beyond the HTS up to 2010.5, the TSP intensity actually decreases on average but still includes large periodic and episodic time variations. Corresponding periods of increased 6-14 MeV electron intensity are observed that have previously been attributed to locally accelerated electrons (McDonald, et al., 2009).

**The Data – The Time Period from ~2007.4-2008.4**

We will consider the temporal variations bracketed by interval ① in Figure 1. This time interval extends from before the HTS crossing to about 2008.4, almost a year after the HTS crossing. This time interval is shown in more detail in Figure 2. Large periodic intensity changes of factors up to ~3 are observed in both the >0.5 and 3.3-8.1 MeV protons plotted here (see Decker, et al., 2008 and Decker, Krimigis, Roelof and Hill, 2010, for time-intensity profiles of lower energy 28-540 KeV ions during this same time period). Some of these increases are also observed for 6-14 electrons, but in many cases the large structured 42/21 day variations are not present for electrons. Much weaker intensity changes are observed for ACR at higher energies (see Figure 1). The higher energy ACR appears to be part of a different population. This is particularly evident in the longer term trends of intensity.

The largest intensity changes of TSP in time interval ① have a period 42-44 days with evidence of a sub-multiple period of ~21-22 days. They are a continuing part of a series of strong periodic intensity changes of TSP extending all the way back to 2007.42, well before V2 first encountered the HTS. In fact, the general increase in both TSP and electron intensities begins at about 2007.55 , 0.1 year (~0.3 AU) before the HTS crossing itself at 2007.66. The sub-multiple period, ~22 days, is strongest just after the HTS crossing and particularly strong during the time period from ~2007.83 to ~2008.0 when the “normalized” 3.3-8.1 MeV rate exceeds the >0.5 MeV rate by a significant amount, as shown by the solid blue data points. The normalized 3.3-8.1 MeV intensity is a factor ~1.5 times higher than the 0.5 MeV rate at 2007.87 (day 318) near the time of a sub-multiple peak. The “normal” value of this ratio is between 0.3-0.6. This enhancement is evidence of a very distorted TSP spectrum with an excess of particles between 3.3-8.1 MeV relative to the lower energy particles at these times.

It is interesting that the 6-14 MeV GCR electron intensity shows a peak (at 2007.97, day 355) at a time ~ 40 days after the 2007.87 peak of 3.3-8.1 MeV protons but at a time when the 3.3-8.1 MeV proton intensity now shows an extreme minimum. These features and their + and – charge asymmetry may provide clues about the acceleration process occurring at this time.

We note that these extreme maxima and minima for electrons and TSP at about 2007.97 occur at the time Intriligator, et al., 2010, report the presence of high energy solar wind ions (HEI) at V2. Burlaga, et al., 2009, also report unusual magnetic field variations (including the onset of a uni-polar region) near the times of the extreme intensity maxima and minima.

High time resolution data for TSP of several energies and for 6-14 MeV electrons is shown in Figure 3 for this time period. Note that the normalized 3.3-8.1/>0.5 MeV ratio for TSP is changing most rapidly during the time period at which the electron intensity is a maximum. Also note that this sudden decrease disappears at higher energies so it is clearly only a TSP effect.

Overall we believe that the time period from about 2007.75 to the end of 2007 represents a period in which the spectrum of TSP below a few MeV is very distorted, possibly indicating that these particles are accelerated in the vicinity of these periodic structures, in close proximity to, but just beyond the HTS. This time period has earlier been identified on a broader time scale of 52 day intervals as one of changing TSP spectra at energies below a few MeV (Cummings, et al., 2009).

The periodic structures that we observe and the acceleration process may be enhanced as a result of the interaction of the interplanetary shock from the large December 2006 events at the Earth with the HTS (Intriligator, et al., 2010). The same process that accelerates the TSP may also accelerate electrons of several MeV. This time period will be discussed in more detail in a later section.

After 2008.0 the >0.5 and 3.1-8.1 MeV TSP intensities continue to decrease but have intensity maxima at 2008.0, 2008.14/2008.20 and 2008.33, which indicate a 44-50 day periodicity with possible submultiples. By 2008.4 the TSP intensities have decreased so that they are comparable to what they were prior to the HTS crossing. The overall electron intensity enhancement also decreases throughout this period, but remains substantial, and does not exhibit the large periodic variations observed for TSP.

**Discussion – The Large Periodic Variations Just Before and After V2 Crossed the HTS**

This part of the discussion concerns the strong 42/21 day periodicities observed in the TSP intensities and the associated rapidly changing TSP energy spectrum which dominate the time period at V2 from about 2007.4, just before the HTS crossing, to about 2008.0. Note that the 1$^{st}$ intensity peak in the ~42 day TSP periodicity occurs at 2007.42 prior to the HTS crossing. This peak is coincident with a B field enhancement prior to the HTS crossing that is described as an MIR by Burlaga, et al., 2009. This time also corresponds to the first of two solar wind plasma speed and ram pressure decreases reported by Richardson, et al., 2008. These decreases prior to the HTS crossing reduce the solar wind ram pressure to ~0.3 of the value it had earlier in 2007 in the upstream solar wind.

The large increases in TSP intensity and the rapid spectral changes at low energies after 2007.66 represent a continuation of this 42 day periodicity after V2 crossed the HTS. The 6-14 MeV electrons also increase rapidly during this time period starting at about 2007.55, which is the time of a second B field enhancement and a second solar wind speed and ram pressure decrease prior to the HTS encounter. The overall intensity maximum for electrons also occurs between about 2007.8-2008.0. The amplitudes of the electron periodicities are weaker than those of the TSP, however, and the electron intensity maxima that do occur (e.g., at 2007.64 and 2007.72) are at the times of the TSP intensity minima (see Figure 2).

After the HTS crossing, the TSP intensities first peak at 2007.77 (day 283) ~41 days after the HTS crossing. During subsequent 21 day sub-cycles the normalized 3.3-8.1/>0.5 MeV ratio increases rapidly up to values ~1.5, from a normal ratio of between 0.3-0.6. This ratio is at a maximum at about 2007.87 (day 318). This increased ratio continues until after the extreme TSP intensity minimum at 2007.97 (at a 21 day sub-cycle minimum, see Figure 3), after which the normalized 3.3-8.1/>0.5 MeV ratio then decreases more slowly over a period of 0.15-0.20 year to the more normal ratio ~0.5. The 1.9-2.7 />0.5 MeV ratio (not shown here) follows a similar intensity-time behavior. It is therefore during the specific time period from about 2007.75-2008.0, with a maximum at 2007.87, that we believe the distortion of the TSP spectrum below a few MeV is at a maximum and the intensity between 3.3-8.1 MeV is enhanced relative to lower energies in alignment with the specific 42/21 day cycles/sub-cycles of maximum intensity. This is a time period in which Cummings, et al., 2009, have observed significant changes in the shape of the TSP spectra in the few MeV range averaged over 26 or 52 day time periods. The striking TSP intensity minimum at 2007.97 coincides with the absolute maximum of 6-14 MeV electron intensities as noted earlier (Figure 3). The intensities of the two oppositely charged components are thus out of phase and the intensity ratio of electrons to protons at roughly the same energy changes by a factor of ~6 in just a few days during this time.

This feature suggests the establishment of a plasma double layer as a possible source of the accelerated TSP. Such double layers have been suggested as the cause of various auroral phenomena in the magnetosphere where an overall acceleration of KeV is possible (Alfren, 1958). In the outer heliosphere these regions may be much larger in scale providing acceleration in the MeV range as observed by V2 (see http://en.wikipedia.org/wiki/Double_layer_(plasma)).

It is interesting to note that Intriligator, et al., 2010, have observed high energy solar wind ions (HEI) at roughly twice the average solar wind speed for a time period between days 333-348 (2007.91-2007.95) almost exactly at the time when we observe a maximum spectral distortion of TSP and a maximum difference in electron and TSP intensities. These energetic solar wind ions have been observed on several other occasions, including at 2004.32 in conjunction with the passage of the “Halloween” shock at V2 and also at 2007.66 when V2 initially crossed the HTS.

This time period near the end of 2007 also includes a period F described by Burlaga, et al., 2009. This time period has two peaks in magnetic field strength reaching values ~0.2 nT, several times the average value, on days 342 and 370 (~28 days separation). Systematic abrupt variations of magnetic field direction from 90° to 270° are observed starting at about 2007.3. They also find a magnetic field with a direction ~0° (directly outward from the Sun) between days 358-367 (in close conjunction with the 6-14 MeV electron increase and the TSP extreme intensity minimum at 2007.97).

The post HTS solar wind “ram” pressure also shows strong peaks at 2007.83 and 2007.96 (~46 d separation) (Richardson, et al., 2008). The ram pressure beyond the HTS at these times approaches the ram pressure measured in the solar wind well upstream of the HTS. These solar wind ram pressure peaks coincide with the strongest intensity maximum and deepest intensity minimum for electrons and TSP respectively.

## The Energy Spectra of the Nuclear and Electron Components

We have averaged the data in time intervals of width ± 3 days centered on the times of 2007.66, 2007.87 and 2008.19 to obtain representative intensities and spectra of TSP and electrons in the above time intervals. To determine the spectra for TSP we have used the energy intervals >0.5 (0.5-1.9), 1.9-2.7, 3.5-5.3, 5.3-7.3, 6.2-10.9, 10.9-16.8 and 16.8-28 MeV. For electrons we have used the energy intervals 2.5-5.3 and 6-14 MeV. The >0.5 MeV intensity in cts/sec (÷10) is plotted as an approximate differential intensity at 1.0 MeV.

The spectra for TSP and electrons for these three time intervals are shown in Figure 4 with the electron intensities multiplied by 100. The intensities are all multiplied by $E^2$ to better illustrate the sometimes subtle spectral differences in the different time intervals. All intensities are plotted at 0.33 of the energy interval width above the lower energy limit to allow for the steep spectra. These effective energies are 1.0, 2.15, 4.0, 6.0, 8.2, 12.8 and 20.5 MeV for TSP and 3.4 and 8.7 MeV for electrons. The TSP spectra at all of these three times appear to steepen with increasing energy with an inflection point at ~2-3 MeV as also reported by Cummings, et al., 2009, 2011. The maxima in the $xE^2$ spectra, or the inflection point of the spectrum, signifies that this energy corresponds to the maximum energy input to the TSP spectrum, an important point for theoretical consideration (see Florinski, et al., 2009).

We have compared our TSP spectrum for the interval 2007.66 with a detailed energy spectrum shown in Stone, et al., 2008, for a 12 day interval near the HTS crossing (see Figure 4). The agreement between spectra is good. The differences at higher energies are due to the fact that the estimated ACR component is subtracted in this paper.

The degree of flattening of the spectral exponent at low energies is determined directly from the changing 3.3-8.1/>0.5 MeV ratio. The larger this "normalized" ratio becomes the greater is the change in the spectral exponent. This 3.3-8.1/>0.5 MeV ratio is larger by a factor of between 2 and 3 for the 2007.87 (A) time period as compared with a later time period at 2008.19 (B). The spectrum for interval A is therefore flatter below ~6 MeV than period B and also flatter than the spectrum at the time of the HTS crossing. It is as though the particles ~1 MeV or less are being removed from the spectrum and re-distributed at energies above ~3.3 MeV during time interval A. Similar spectral features have been reported by Cummings, et al., 2006, before and after the HTS crossing by V1 in late 2004.

In the energy range above 5 MeV/nuc the TSP have spectra with exponents in the range ~3.1-3.6. The spectral exponent in Period A, when there is believed to be acceleration, appears to be slightly flatter (3.20±0.10) than the spectrum in period B (3.60±0.15) when no acceleration appears to be occurring. The errors on the “differences” in spectra are mainly statistical since the energy intervals are identical for each time interval. The statistical errors are less than a few percent for all TSP intervals.

For electrons in the energy range from 3.4-8.7 MeV, the spectral exponents are also in the range -3.2-3.6 for each time period, albeit with larger errors. So electrons and TSP have very similar spectral exponents in this energy range. The ratio of TSP intensities to electron intensities in this energy range is ~500 ±150 with the errors representing the variability in these ratios from time period to time period, and also from lower to higher energies.

## Summary and Conclusions

This paper summarizes high time resolution data on TSP intensities between ~0.5 and 20 MeV and electrons from 6-14 MeV obtained by the CRS experiment on V2 in 2007 before and after this spacecraft crossed the HTS.

Large periodic intensity variations dominate the intensity-time landscape of TSP and electrons just before and after the HTS crossing at 2007.66. These variations provide evidence of a highly structured and variable heliosheath at V2 near the HTS. Much of the variability in the TSP and electron intensities is in the region from just in front of the HTS to perhaps ~1-2 AU beyond the HTS.

The strong 42/21 day periodicities of TSP intensity observed just before and just after V2 encountered the HTS are perhaps the most striking aspect of these variations. The largest amplitude periodicities are observed at energies between about 0.5 and 10 MeV, but not so much at still lower energies, where the intensity variation at these times is much less in the LECP experiment, (Decker, et al., 2010 and http://voyager.gsfc.nasa.gov). Intensity changes greater than a factor ~2 are observed between the intensity maximum and minimum in any one period and changes in the TSP spectral index below a few MeV are evident at times of both the intensity maxima and minima during the time interval from about 2007.75 to 2008.0. These spectral changes are evidenced by the changes in the 3.3-8.1/>0.5 MeV ratio which acts as an index of the spectral changes. An increasing ratio indicates a more rapid intensity increase of higher energy TSP relative to the lowest energy (which itself is increasing) during this time period, suggestive of acceleration of several MeV TSP near the energy peak of the x $E^2$ spectrum (which occurs at ~5 MeV at this time) by a mechanism related to the specific 42/21 day periodicities themselves.

The electron periodicities during this time interval are not as striking as those of TSP. The electron intensities are well above background so the electrons appear to be locally accelerated, (e.g., McDonald, et al., 2009). The electron intensity peaks that do occur are out of phase with the 21 day intensity sub peaks of the TSP, however. The largest excess of 6-14 MeV electrons, in fact, occurs at the deepest minimum in the intensity of TSP at 2007.97.

This intense activity and the periodicities in the intensity at this time may be instigated by the arrival of a large IP shock at the HTS. This could be the interplanetary shock related to the December 2006 events at the Earth. This shock propagates out to the V2 location (Intriligator, et al., 2010; Webber and Intriligator, 2011) arriving in the later half of 2007. This may at least partially account for the much larger intensity of TSP observed when V2 crossed the HTS in 2007 relative to when V1 crossed the HTS 2.7 years earlier and also for the fact that the TSP intensities at V2 peaked in the heliosheath region just beyond the HTS in 2007 whereas at V1 the TSP intensities steadily increased after crossing the HTS and peaked much deeper into the heliosheath.

If the two cosmic ray/B field/solar wind plasma "events" that were observed by V2 at ~2007.42 and 2007.55, described as MIR's by Burlaga, et al., 2009, are attributed to fore-structures of the HTS itself; then the December, 2006 shock would have had to have reached the HTS and therefore V2, not prior to the HTS crossing, but sometime in the later part of 2007 perhaps between about 2007.7 and 2008.0. This is consistent with the propagation time of this shock as described by Webber and Intriligator, 2011.

We have obtained both TSP and electron energy spectra for three time periods during this very disturbed period including the time of the HTS crossing. The TSP and electron spectra in the energy range from ~3 MeV up to ~20 MeV are similar with average spectral exponents in the range ~3.0-3.6. These spectral exponents are close to, but perhaps slightly larger than the canonical value of 3.0 expected for diffusive shock acceleration at a strong shock, where s=2.0.

The typical TSP/electron intensity ratio during the averaging periods is ~500 with a large variability.

The TSP energy spectra flatten significantly below ~3 MeV as shown in the x $E^2$ spectral plots. This is most notable in the x $E^2$ spectrum for time interval A which has a higher peak intensity between 3-6 MeV than interval B or the HTS interval, but the intensities in time interval A are the same or less than the other two intervals at energies ~1 MeV or below. This degree of flattening is greatest when the normalized 3.3-8.1/>0.5 MeV ratio is largest and this implies that particles ~1 MeV or less are removed from the spectra at these times and possibly redistributed at energies peaking at ~5 MeV where the peak in the x $E^2$ spectrum is observed.

Overall this study emphasizes the strong 42/21 day periodicities observed in the intensities of 1-10 MeV TSP just before and after V2 crossed the HTS in 2007. These periodicities were accompanied by changes in the spectrum of TSP below ~3 MeV along with the acceleration of 6-14 MeV electrons. The intensity peaks of electrons are generally ~21 days out of phase with the intensity peaks of TSP. We believe that these features are evidence for significant acceleration of both TSP and electrons during these time periods. The difference in phase of the electrons and protons suggests that a plasma double layer capable of accelerating particles to MeV energies may be a part of this acceleration process.

**Acknowledgements:** The authors all appreciate the support of the Voyager program by JPL. The data used here comes mainly from internally generated data bases which may be accessed via e-mail through the web-site (http://voyager.gsfc.nasa.gov).

## FIGURE CAPTIONS

**Figure 1:** Intensities of TSP (>0.5 MeV) and 6-14 MeV electrons (red) at V2 from 2007.3 to 2011.3. Data are 5 day running averages. Intensities are $cts \cdot s^{-1}$ for >0.5 MeV and $P/m^2 \cdot s \cdot sr \cdot MeV$ (x 100) for electrons. Time periods of interest are labeled 1 and 2. Note that the vertical scale is linear and greatly expanded. The solid red line is the estimated background (both GCR and instrumental) for electrons. The relative intensities of 27-42 MeV He nuclei as representative of the anomalous component are also shown in blue.

**Figure 2:** The intensities during time interval 1 at V2. Intensity scale for >0.5 MeV is in $cts \cdot s^{-1}$. The 3.3-8.1 MeV TSP (blue) and 6-14 MeV electron intensities (red) are "normalized" by the factors indicated and the units of intensity are $P/cm^2 \cdot s \cdot sr \cdot MeV$. 6-14 MeV electrons are in red and are the "excess" electron intensities above background. The vertical lines indicate the times of prominent peaks for >0.5 MeV particles. Upper numbers between the lines are in days. Lower numbers refer to the sub-multiple peaks. Note that the times when the normalized 3.3-8.1 MeV intensity exceeds >0.5 MeV intensity are shown by the solid blue data points.

**Figure 3:** Higher time resolution data showing several TSP and electron intensities during the time interval from 2007.9 to 2008.04.

**Figure 4:** Energy spectra of TSP and electrons (intensities x $E^2$ in MeV) observed during the time intervals A and B and also at the time of the HTS crossing (electron spectra in red and intensities x $10^2$). Detailed 12 day average energy spectra from Stone, et al., 2008, for the HTS interval are also shown (in blue). Further details regarding these spectra are given in the text.

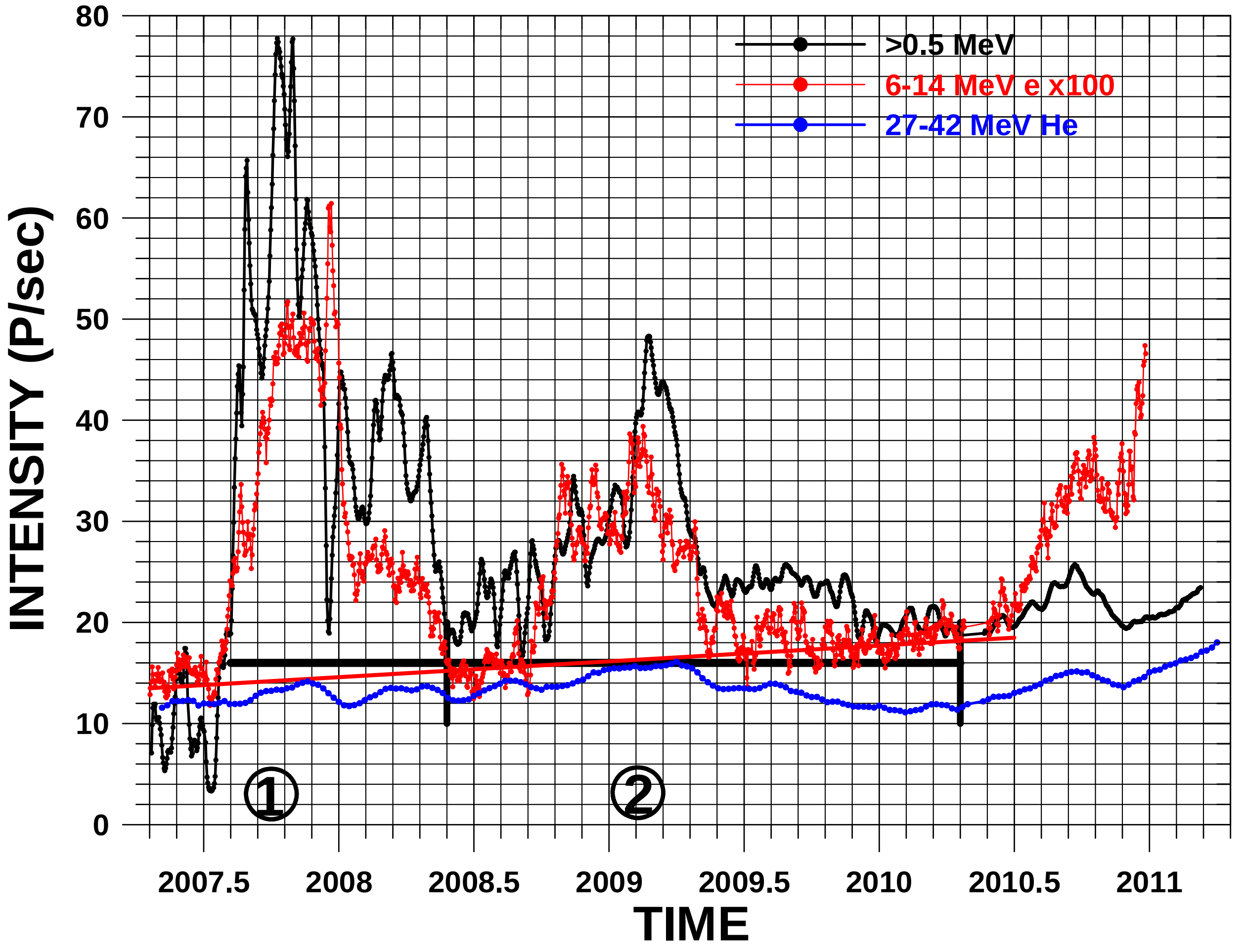

FIGURE 1

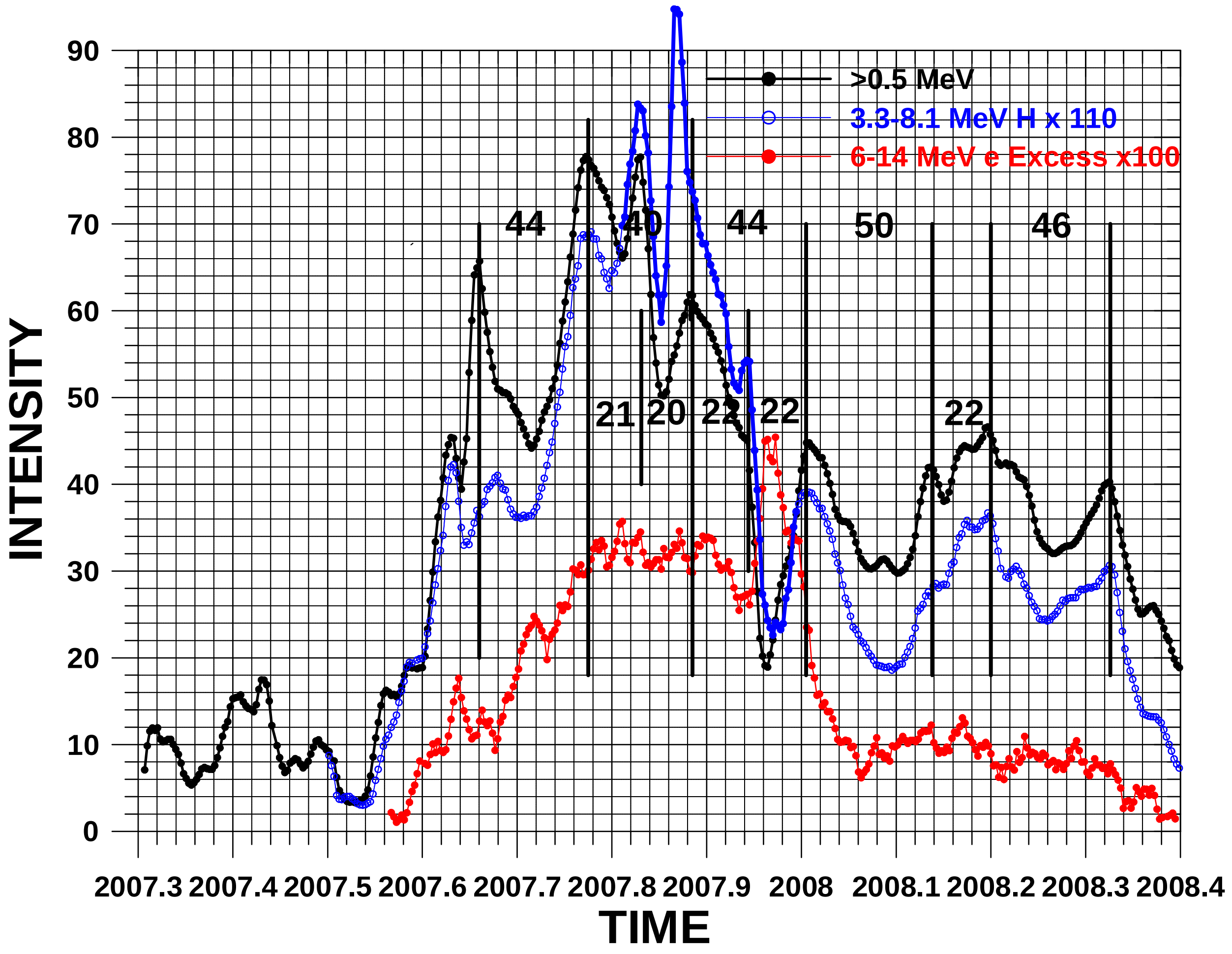

>0.5 MeV
3.3-8.1 MeV H x 110
6-14 MeV e Excess x100
INTENSITY
0
10
20
30
40
50
60
70
80
90
44
40
44
50
46
21
20
22
22
22
2007.3 2007.4 2007.5 2007.6 2007.7 2007.8 2007.9 2008 2008.1 2008.2 2008.3 2008.4
TIME

**FIGURE 2**

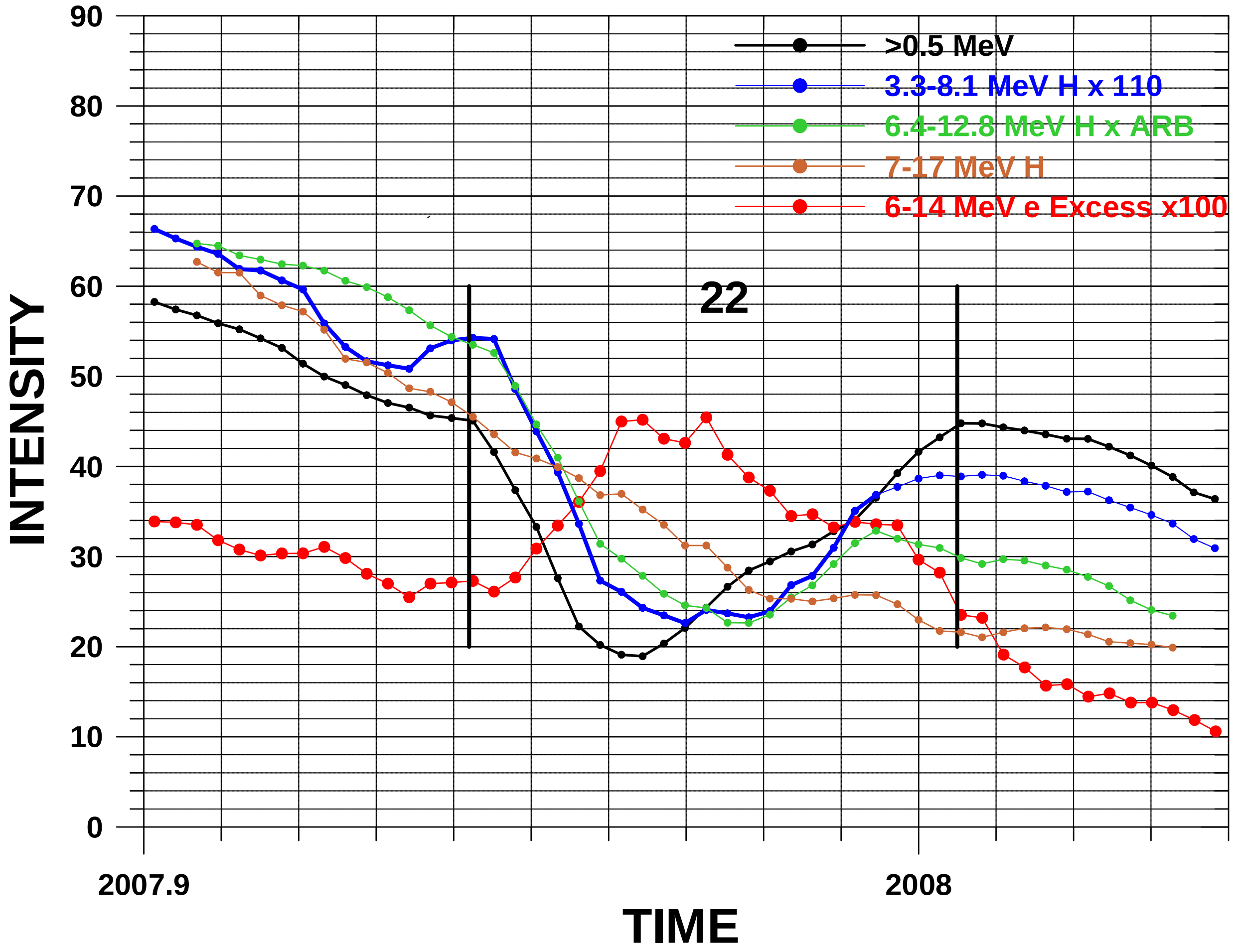

FIGURE 3

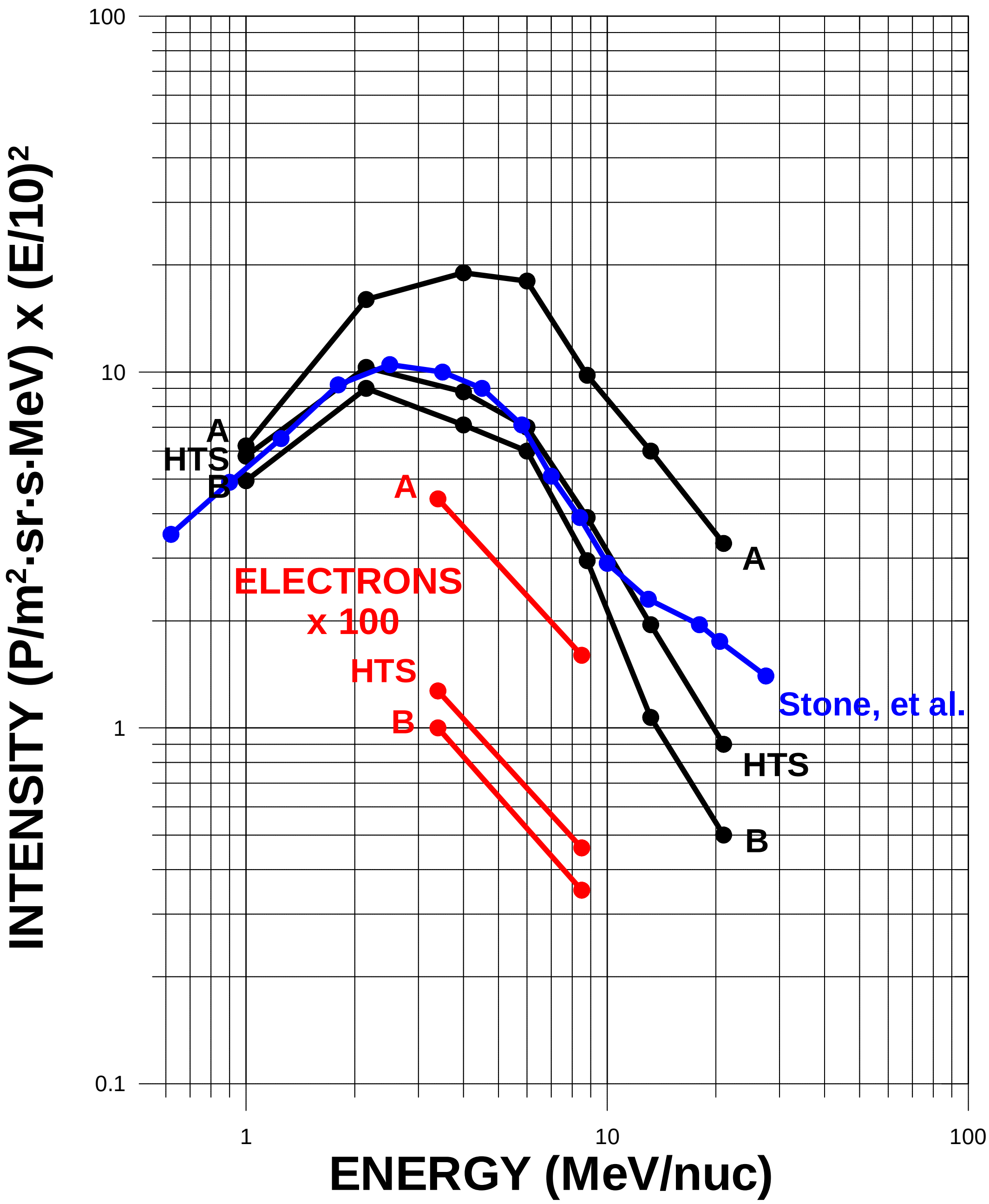

FIGURE 4